# Self-Reported Side Effects of Semaglutide and Tirzepatide in Online Communities

Neil K. R. Sehgal, ME[1,2], Jena Shaw Tronieri, PhD[3], Lyle Ungar, PhD[1], Sharath Chandra Guntuku, PhD[1,2]

[1]Computer and Information Science Department, University of Pennsylvania, Philadelphia, PA, USA
[2]Leonard Davis Institute of Health Economics, University of Pennsylvania, Philadelphia, PA, USA
[3]Department of Psychiatry, Perelman School of Medicine, University of Pennsylvania, Philadelphia, PA, USA

**ABSTRACT**

Social media can reveal patient experiences with glucagon-like peptide-1 receptor agonists (GLP-1 RAs) that extend beyond clinical trial data. We analyzed 410,198 Reddit posts (May 2019–June 2025) mentioning semaglutide or tirzepatide. A total of 67,008 users self-reported using these medications, and 43.5% described at least one side effect. Gastrointestinal symptoms predominated, including nausea (36.9%), fatigue (16.7%), vomiting (16.3%), constipation (15.3%), and diarrhea (12.6%). Notably, reproductive symptoms (e.g., menstrual irregularities) and temperature-related complaints (e.g., chills, hot flashes) emerged as unrecognized potential effects. These findings highlight patient concerns not well captured in current labeling or trials. Large-scale social media analysis can complement traditional pharmacovigilance by detecting emerging safety signals and expanding understanding of the real-world safety profile of GLP-1 RAs.

**MAIN TEXT**

Two novel glucagon-like peptide-1 receptor agonists (GLP-1 RAs), semaglutide and tirzepatide, have rapidly gained popularity for type 2 diabetes management and weight control. Clinical trials and regulatory documents have established gastrointestinal symptoms, such as nausea, vomiting, and diarrhea, as common side effects of these medications. Social media data provide the opportunity to capture a broader range of patient experiences, including symptoms not formally recognized in drug labeling.[1] Understanding these real-world reports can help clinicians and regulators build a more comprehensive safety profile.

In this paper, we systematically characterize self-reported side effects of semaglutide and tirzepatide shared on Reddit, a social media platform with over 100 million daily active users anonymously posting in topical forums (subreddits).[2] Reddit was selected as the data source for this study due to its large, publicly accessible, topic-specific health communities and its long-form, context-rich user discussions. Prior work has established Reddit as a valuable source for studying health-related disclosures, including sensitive or stigmatized medical experiences, and as an important resource for public health and computational social science research.[3,4] Unlike other major platforms, which provide either restricted, unreliable, or prohibitively costly API access, publicly released researcher-curated corpora have made Reddit data practically feasible for large-scale public health research.

We conducted a cross-sectional study of self-reported side effects of semaglutide and tirzepatide using Reddit data from May 2019 through June 2025. Posts/comments were collected from nine large subreddits that discuss GLP-1 RAs or weight management.

We first identified posts where users disclosed taking semaglutide or tirzepatide in any Food and Drug Administration (FDA) approved formulation for type 2 diabetes or weight management using a GPT-4o-mini based classifier. This classifier also extracted the specific medications mentioned. Among posts indicating self-use, we applied a GPT-4.1-mini based classifier to extract self-reported side effects and map them to MedDRA Preferred Terms (PTs).

Across the May 2019–June 2025 period during which eligible posts appeared, 172,679 of 410,198 Reddit posts mentioning semaglutide or tirzepatide indicated personal use by 67,008 users. Among them, 43.5% (29,172 users) reported at least one side effect, averaging 2.7 (SD 2.4) PTs per user.

Table 1 presents the prevalence of PTs among users reporting side effects. Nausea was most common (36.9%), followed by fatigue (16.7%), vomiting (16.3%), constipation (15.3%), and diarrhea (12.6%). Other symptoms reported by ≥5% of users who disclosed side effects included decreased appetite, abdominal pain, gastroesophageal reflux, headache, abdominal distension, and dizziness. The most frequent co-occurring PTs were nausea and vomiting, with 2,917 users (10.0%) self-reporting both (Appendix Table 1).

**Table 1. Frequency of MedDRA Preferred Terms (PTs) among 29,172 Reddit Users Taking GLP-1 RAs**

PTs are grouped by primary System Organ Classes (SOCs) (bolded) in descending order. PTs reported by fewer than 0.5% of users and SOCs reported by fewer than 1.0% are omitted for brevity. See Appendix Table 3 for PTs ordered purely by frequency, and Appendix Tables 4–5 for frequencies of corresponding MedDRA High Level Terms (HLTs) and High Level Group Terms (HLGTs).

| MedDRA Term | # Users Self Reporting | % of Users with Self-Disclosed PT |
|---|---:|---:|
| **Gastrointestinal disorders** | 19185 | 65.8 |
| Nausea | 10764 | 36.9 |
| Vomiting | 4749 | 16.3 |
| Constipation | 4463 | 15.3 |
| Diarrhea | 3686 | 12.6 |
| Abdominal pain | 2493 | 8.5 |
| Eructation | 2010 | 6.9 |
| Gastroesophageal reflux disease | 1877 | 6.4 |
| Abdominal distension | 1488 | 5.1 |
| Flatulence | 1055 | 3.6 |
| Abdominal discomfort | 900 | 3.1 |
| Gastrointestinal disorder | 883 | 3.0 |
| Dyspepsia | 662 | 2.3 |
| Retching | 350 | 1.2 |
| Dry mouth | 321 | 1.1 |
| Breath odour | 187 | 0.6 |
| Gastrointestinal motility disorder | 167 | 0.6 |
| Impaired gastric emptying | 167 | 0.6 |
| Dysphagia | 142 | 0.5 |
| **General disorders and administration site conditions** | 8395 | **28.8** |
| Fatigue | 4870 | 16.7 |
| Asthenia | 738 | 2.5 |
| Early satiety | 666 | 2.3 |
| Pain | 619 | 2.1 |
| Malaise | 589 | 2.0 |
| Chills | 332 | 1.1 |
| Injection site pain | 283 | 1.0 |
| Hunger | 220 | 0.8 |
| Feeling cold | 183 | 0.6 |
| Illness | 156 | 0.5 |
| Thirst | 147 | 0.5 |
| Injection site reaction | 144 | 0.5 |
| Pyrexia | 139 | 0.5 |
| **Nervous system disorders** | 5704 | **19.6** |
| Headache | 1784 | 6.1 |
| Dizziness | 1445 | 5.0 |
| Somnolence | 492 | 1.7 |
| Dysgeusia | 398 | 1.4 |
| Syncope | 346 | 1.2 |
| Tremor | 333 | 1.1 |

| | | | |
|---|---:|---|---:|
| Lethargy | 311 | | 1.1 |
| Paraesthesia | 301 | | 1.0 |
| Migraine | 294 | | 1.0 |
| Cognitive disorder | 267 | | 0.9 |
| Burning sensation | 190 | | 0.7 |
| Taste disorder | 177 | | 0.6 |
| **Metabolism and nutrition disorders** | 5205 | | 17.8 |
| Decreased appetite | 3371 | | 11.6 |
| Hypoglycemia | 600 | | 2.1 |
| Increased appetite | 549 | | 1.9 |
| Dehydration | 473 | | 1.6 |
| Food aversion | 365 | | 1.3 |
| Food craving | 171 | | 0.6 |
| **Psychiatric disorders** | 3756 | | 12.9 |
| Anxiety | 1230 | | 4.2 |
| Insomnia | 901 | | 3.1 |
| Depression | 827 | | 2.8 |
| Panic attack | 280 | | 1.0 |
| Irritability | 257 | | 0.9 |
| Anhedonia | 191 | | 0.7 |
| Libido decreased | 161 | | 0.6 |
| Mood swings | 159 | | 0.5 |
| Apathy | 149 | | 0.5 |
| Sleep disorder | 148 | | 0.5 |
| Depressed mood | 142 | | 0.5 |
| **Skin and subcutaneous tissue disorders** | 2552 | | 8.7 |
| Alopecia | 914 | | 3.1 |
| Hyperhidrosis | 362 | | 1.2 |
| Pruritus | 335 | | 1.1 |
| Rash | 193 | | 0.7 |
| Acne | 183 | | 0.6 |
| Dermatitis allergic | 166 | | 0.6 |
| Urticaria | 149 | | 0.5 |
| **Musculoskeletal and connective tissue disorders** | 1833 | | 6.3 |
| Myalgia | 540 | | 1.9 |
| Arthralgia | 336 | | 1.2 |
| Back pain | 325 | | 1.1 |
| Muscle atrophy | 194 | | 0.7 |
| Muscle spasms | 158 | | 0.5 |
| Musculoskeletal pain | 134 | | 0.5 |
| **Reproductive system and breast disorders** | 1142 | | 3.8 |
| Intermenstrual bleeding | 266 | | 0.9 |
| Heavy menstrual bleeding | 261 | | 0.9 |
| Menstruation irregular | 211 | | 0.7 |
| Menstrual disorder | 178 | | 0.6 |
| Dysmenorrhoea | 168 | | 0.6 |
| **Investigations** | 832 | | 2.9 |

| | | |
|---|---:|---:|
| Weight increased | 415 | 1.4 |
| **Cardiac disorders** | 718 | 2.5 |
| Tachycardia | 512 | 1.8 |
| Palpitations | 205 | 0.7 |
| **Respiratory, thoracic and mediastinal disorders** | 673 | 2.3 |
| Dyspnoea | 169 | 0.6 |
| **Vascular disorders** | 628 | 2.2 |
| Hypotension | 166 | 0.6 |
| Hot flush | 151 | 0.5 |
| **Infections and infestations** | 453 | 1.6 |
| **Eye disorders** | 388 | 1.3 |
| Visual impairment | 218 | 0.7 |
| **Renal and urinary disorders** | 357 | 1.2 |
| **Injury, poisoning and procedural complications** | 234 | 0.8 |
| Contusion | 157 | 0.5 |

Overall, 65.8% of semaglutide/tirzepatide users with side effects reported at least one gastrointestinal symptom. Psychiatric symptoms were noted by 12.9%, with anxiety (4.2%) and depression (2.8%) most common.

As a secondary analysis, we examined monthly frequencies of the ten most commonly reported MedDRA Preferred Terms. Symptom counts increased in parallel with overall discussion volume but did not show distinct or interpretable temporal shifts (Appendix Figure 1).

As a secondary analysis, we conducted a sub-analysis of users who exclusively mentioned one formulation to determine differential side effect rates. Of the 29,172 users who self-disclosed at least one side effect, 17,937 (61.5%) exclusively mentioned semaglutide formulations, and 7,125 (24.4%) exclusively mentioned tirzepatide formulations.

In exploratory descriptive analyses, we examined side effect frequencies separately for users who exclusively mentioned semaglutide (n=17,937) versus tirzepatide (n=7,125). Among semaglutide users reporting side effects, the most commonly reported symptoms were nausea (39.4%), fatigue (16.1%), vomiting (18.0%), constipation (14.9%), and diarrhea (12.4%). Among tirzepatide users reporting side effects, the most commonly reported symptoms were nausea (28.6%), fatigue (14.7%), vomiting (11.1%), constipation (12.9%), and diarrhea (12.5%). Injection site reactions, myalgia, insomnia, and temperature-related symptoms (chills, feeling cold) were each reported by 1-4% of tirzepatide users. A complete descriptive summary for each formulation is provided in Appendix Table 2.

In this large-scale Reddit analysis, over 67,000 semaglutide or tirzepatide users were identified, with 44% reporting side effects. Consistent with the findings of clinical trials, gastrointestinal symptoms were reported most commonly by these users.[5-8] The most frequently reported gastrointestinal effects were nausea, vomiting, constipation, diarrhea, and abdominal pain, which are the only adverse reactions featured as occurring more commonly than with placebo in the FDA-approved prescribing information for all five versions of these medications. Additional symptoms reported by ≥5% of users (e.g., decreased appetite, eructation, distension, dizziness) have similarly been noted in trials as more common than placebo.[5-8] Headache, though common in trials, shows mixed placebo comparisons, while fatigue, Reddit's second most reported symptom, has met reporting thresholds in relatively few trials.[5-8]

We also identified several side effects not previously reported with semaglutide or tirzepatide. Nearly 4% of Reddit users with side effects reported reproductive symptoms, notably menstrual changes like intermenstrual bleeding (0.9%), heavy bleeding (0.9%), and irregular cycles (0.7%). These rates would be higher in female-only samples. While weight loss may alleviate some reproductive issues, both weight gain and loss are linked to menstrual irregularities in women with obesity.[9] Further, GLP-1 RAs are thought to impact food intake, in part, by engaging receptors in the hypothalamus, a brain region that also plays a central role in menstrual cycle regulation.[10] Additionally, some users described symptoms possibly linked to altered temperature regulation (e.g., chills, hot flashes, pyrexia), warranting further investigation in placebo-controlled trials or post-market data, particularly due to glucagon's associations with increased thermogenesis.[10]

Strengths of our study include a large and contemporary sample, the focus on unprompted, self-disclosed experiences, the systematic classification of symptoms using MedDRA, and the development of a straightforward LLM-based pipeline to detect emerging, potentially under-reported signals in a timely manner.

Several limitations should be noted. First, patients using weight loss subreddits may differ from the overall population prescribed GLP-1 RAs. Specifically, Reddit users tend to be younger, are more likely to be male, and are disproportionately located in the United States.[11,12] The demographic composition of Reddit users and self-selection into health-related discussions limit the generalizability of our findings. Without cross-platform validation or comparison to structured pharmacovigilance databases, we cannot quantify the magnitude or direction of these biases. Although Reddit provides one of the largest and most active online health communities and remains among the few platforms with accessible, high-volume data, future work should extend this approach to other sources–such as specialized patient forums, clinical review platforms, or any social media systems that restore stable API access–to assess the consistency of patient-reported symptom patterns across environments. Our results should be interpreted as hypothesis-generating signals that require confirmation through traditional surveillance systems and prospective studies in representative populations.[13]

Second, because users were not prompted to disclose all side effects, we cannot estimate their true prevalence. Participation in health-related subreddits is voluntary, and individuals who experience more severe, persistent, or distressing symptoms may be more likely to post about them than those with neutral or positive experiences. Even those who reported side effects may not have disclosed every symptom experienced, and individuals' beliefs about which symptoms are potentially attributable to the medication may have influenced their report. Further, prior work has found that while clinical trials often rely on self-report, gastrointestinal symptoms are often under-reported and inconsistently reported compared to validated measures for functional

GI disorders.[14,15] For these reasons, self-reported experiences cannot be used to determine true event frequency or causality. The symptoms discussed here may be attributable to weight loss itself, changes in comorbidities, or other unrelated factors. Additionally, dosage, route of administration, and treatment duration were inconsistently reported in posts and were not analyzed. Our analysis prioritized breadth of symptom identification over temporal granularity. While we examined monthly trends in common symptoms (Appendix Figure 1), detailed assessment of symptom onset timing or dose-response relationships would require structured metadata (e.g. prescription dates, dose titration schedules) rarely disclosed in social media posts. Examination of these variables would require data sources with more complete and structured medication histories, such as electronic health records or prospective registries. We were unable to determine when or how frequently these events occurred over the course of treatment or whether event occurrence depended on patient characteristics, such as the presence of type 2 diabetes. Results should be viewed as hypothesis-generating signals about patient-perceived symptoms rather than estimates of causal effects or true event rates.

Third, natural language processing can misclassify or overlook nuanced context, and adverse drug event extraction from social media texts remains an active area of research.[16] We validated the specific models deployed in this study against manually annotated samples and achieved high performance comparable to prior adverse event extraction systems[1,16]. While systematic benchmarking across LLM architectures represents an important direction for future methods research, our validation approach ensured reliable extraction for this application. Additionally, certain self-reported effects may not fit neatly into existing MedDRA concepts or may have been misclassified based on the user's description. For example, some events classified as "gastroesophageal reflux" based on user terminology might have instead been recorded as "dyspepsia" in the context of a clinical trial. In our exploratory sub-analysis restricted to users whose posts exclusively mentioned one formulation, we present descriptive comparisons of side effect frequencies but do not make formal statistical inferences about medication-specific effects. Without demographic and clinical data to match or adjust for confounding between medication groups, we cannot determine whether observed differences in complaint frequencies are attributable to the formulations themselves versus differences in the populations using each formulation. Such analyses would require data sources with detailed patient characteristics to enable appropriate comparison groups. Future work should stratify symptom patterns by medication using data sources with richer clinical context and validated methods for establishing medication-specific causal effects.

Despite these limitations, social media data provide important insight into symptoms that are of concern to semaglutide and tirzepatide users and may be helpful in identifying rarer or under-reported effects that warrant further investigation. Online health communities are increasingly influential in shaping patient expectations and treatment decisions, and a growing body of research examines these platforms across a range of contexts.[17-21] Future work could extend this framework by examining more detailed temporal dynamics in symptom patterns and how they vary by medication type or formulation; incorporate severity distinctions in patient-reported symptoms; and evaluate multilingual or non-Reddit digital communities to understand cross-platform and cross-cultural variation in patient experiences.

Capturing these patient-reported experiences can alert clinicians and researchers to emerging patient concerns related to GLP-1 RA use. Our data suggest that clinicians may encounter patient inquiries about menstrual irregularities and body temperature fluctuations in addition to commonly reported gastrointestinal symptoms. We recommend that future prospective trials and post-market surveillance studies systematically assess fatigue and reproductive symptoms, particularly menstrual changes in female participants, using validated instruments rather than relying solely on passive adverse event reporting. These findings may also inform updates to patient counseling materials and regulatory labeling to better address symptoms of concern to patients in real-world settings. We encourage researchers to examine whether these patient-reported concerns occur more frequently with semaglutide and/or tirzepatide than with placebo.

**COMPETING INTERESTS**

JST reports receiving an investigator-initiated grant, on behalf of the University of Pennsylvania, from Novo Nordisk and receiving consulting fees from Currax Pharmaceuticals, LLC.

**METHODS**

Posts/comments were collected from nine large subreddits (forums) that discuss GLP-1 RAs or weight management. These subreddits were selected to align with those analyzed by prior researchers who estimated real-world weight loss outcomes of these medications.[1] Relevant posts and comments containing mentions of semaglutide or tirzepatide were identified using a list of brand and generic names and common misspellings. Data were collected for the period January 1, 2015, through June 30, 2025; in practice, all posts that met inclusion criteria were dated May 2019–June 2025. All data were collected using Pushshift and Arctic Shift (an extension of Pushshift).[2,3] Please see the supplementary methods for subreddits used and a full list of search terms.

We first identified posts in which users personally disclosed taking either semaglutide or tirzepatide in any formulation currently approved by the Food and Drug Administration (FDA) for the treatment of type 2 diabetes or weight management using a GPT-4o-mini based text classifier (prompt in supplementary methods). The classifier also extracted the specific medications being taken. The classifier's self-use performance was evaluated against a manually annotated set of 100 randomly selected posts and had a sensitivity of 86% and specificity of 96%. Only posts that met the self-disclosure criterion were included in subsequent analyses.

We evaluated the model's medication extraction accuracy on a random sample of 100 posts containing 109 medication mentions. The model achieved a precision of 95% and recall of 98% (F1=96%) at the medication level. Common errors included extracting a specific medication when the post did not clearly state which drug was being used, often inferring the medication from the subreddit name (e.g., assuming Ozempic for posts in r/Ozempic) rather than from explicit mentions in the post content (n=4), and missing secondary (non-semaglutide or -tirzepatide) medications (n=2). At the post level, 93.0% of posts had all medications correctly extracted.

Among posts indicating self-use, we applied a second classifier, based on the GPT-4.1-mini model, to extract self-reported side effects. Prior work has shown that both classical and transformer-based NLP systems can map free text to standardized vocabularies such as

MedDRA.[4,5] Our pipeline follows this normalization paradigm but uses a large language model classifier to directly map side effects mentioned in the user-generated text to standardized MedDRA Preferred Terms (PTs) using retrieval augmented generation (prompt in supplementary methods). Manual evaluation of 100 randomly selected posts indicated that the classifier had a sensitivity (recall) of 97% and a positive predictive value (precision) of 87% for identifying the presence or absence of individual side effects. Because true negatives cannot be enumerated meaningfully for named entity recognition models at the span level, specificity is undefined for this granularity. The classifier extracted 2,013 unique side effect phrases, which were matched to existing MedDRA PTs via exact matching and manual review when needed.

Percentages were calculated based on the population of users who disclosed at least one side effect. Because users were not prompted to disclose side effects, we cannot estimate their true prevalence. Further, uncontrolled self-reported experiences cannot be used to determine causality. The symptoms discussed here may be attributable to weight loss itself, changes in comorbidities, or other unrelated factors.

For users exclusively mentioning one formulation (either semaglutide or tirzepatide but not both), we conducted exploratory descriptive comparisons of side effect frequencies. For each MedDRA Preferred Term, we calculated proportions in each medication group to characterize the distribution of reported symptoms. These comparisons are presented descriptively without formal statistical testing, as we lack the demographic and clinical data necessary to adjust for potential confounding factors that may differ between users of each medication.

GPT-4o-mini was chosen for self-disclosure classification because, at the time that component of the study was conducted, it was among the most capable lightweight models available for large-scale processing at feasible cost. Midway through the project, OpenAI released GPT-4.1-mini with documented performance improvements over GPT-4o-mini across a range of tasks including instruction following[6], and we therefore used GPT-4.1-mini for identifying side-effect mentions and mapping them to MedDRA terms. Larger reasoning or non-mini models (e.g., GPT-4o, GPT-4.1) were not employed for cost reasons.

All analyses were conducted in Python 3.11 using standard data processing libraries. Data were collected from publicly accessible Reddit posts and comments using Pushshift and Arctic Shift during the study period. No attempts were made to access private or password-protected content. As data were public and non-identifiable, IRB review was not required. Human Ethics and Consent to Participate declarations: not applicable.

**DATA AVAILABILITY**
All data used in this study are publicly available from Reddit.com, or are accessible via Pushshift and Arctic Shift. Due to Reddit's Terms of Use, we are unable to share the raw data. All Reddit data in this study were used in accordance of Reddit's Terms of Use.

**CODE AVAILABILITY**
Code to reproduce the analyses can be found at https://github.com/sehgal-neil/glp1-side-effects-analysis.

**METHODS ONLY REFERENCES**